# A Multifuntional Frequency-selective Polarization Converter for broadband backward-scattering reduction

LingLing Wang,[1,2] Shaobin Liu,[1,*] Xiangkun Kong,[1] Haifeng Zhang,[1,2] Yongdiao Wen[1,2] and Qiming Yu[1]

[1]*College of Electronic and information Engineering, Nanjing University of Aeronautics and Astronautics, Nanjing, 210016, China*
[2]*College of Electronics and Optical Engineering & College of Microelectronics, Nanjing University of Posts and Telecommunications, Nanjing, 210023, China*

**Abstract:** In this paper, we combine the design of band-pass frequency selective surfaces (FSSs) with polarization converters to realize a broadband frequency-selective polarization converter (FSPC) with low-backward scattering, which consists of the top polarization conversion layer backed by a multi-layer band-pass FSS. It is numerically demonstrated that the 1 dB transmission window can be obtained from 8.5 GHz to 11 GHz with a 25.6% fractional bandwidth (FBW), and the bandwidth of reflection below -10 dB is up to 92% from 5.6 GHz to 15.13 GHz. Moreover, the proposed device can achieve two polarization conversion bands (5.66-6.9 GHz and 12.8-15.2GHz) with the polarization conversion ratio over 90%. Besides, by arranging the proposed structure in a checkerboard-like distribution, the backward scattering energy can be reduced in a wide frequency band ranging from 4 to 16 GHz. Both simulation and experimental results are in good agreements, which demonstrates our design strategy. Compared with the conventional polarization conversion designs, the proposed design presents an extra frequency-selective performance and hence can be applied to various practical situations, for instance, working as radomes to transmit the in-band signals with high-efficiency while keeping low-backward scattering for the out-of-band waves.

## 1. Introduction

Frequency selective surfaces (FSS) usually composed of periodically arranged elements have always been a research hotspot for it exhibits band-pass or band-stop responses [1-3]. Frequency selective radomes, as an important application of band-pass FSSs, can achieve good transmission coefficients in the passband [4-6]. However, the strong reflection out of the transmission window which increases the radar cross section (RCS) greatly becomes a major obstacle for band-pass FSS [7]. To reduce these reflections, researchers have presented a new concept, namely frequency selective rasorber (FSR), which is transparent to incident EM waves in the passband and absorptive outside the passband. FSRs can absorb the EM waves outside the transmission window to reduce the RCS by introducing a lossy layer which consists of lumped resistances or resistive surfaces in front of band-pass FSS [8-10]. However, the use of lumped resistances increases the manufacturing costs because of the welding of high-frequency resistances. In addition, the method of absorbing EM wave by EM-absorbing materials to reduce RCS will dissipate the incoming EM energy into heat which will increase the possibility of being detected by infrared detectors.

Besides the EM-absorbing material to absorb the EM energy, phase cancellation with coding metasurface to dissipate the incoming wave into other directions is an alternative approach to reduce the backward RCS [11-14]. For instance, Li et.al proposed a two-dimensional phase gradient metasurface for wide-band, polarization-independent and high-efficiency RCS reduction [15]. Qiu et.al combine the Pancharatnam-Berry phase metasurface with carbon film to propose a composite metasurface with diffuse scattering and absorbing characteristics to reduce the RCS of a metal target in broadband [16]. Although, the fruitful progress of metasurface to reduce the RCS, most above mentioned designs were confined to mono-functional operations. Recently, Sima et.al reported a bi-functional frequency-selective coding meta-mirror, which combines broadband backward-scattering reduction and high-efficient specular reflection by a single meta-mirror, to perform desired scattering properties in different frequency bands by tailoring the phase dispersion of the metasurface [17].

Owing to the polarization conversion surface that can acquire the 180º phase difference easily by rotating the unit cell 90º degrees, they are one of the options to reduce the backward scattering [18-20]. There are a variety of anisotropic structure can be used to design the polarization converter, including split-ring resonators (SRR) [21, 22], double-head arrow structure [23, 24], short V-shaped wire [25], and fishbone-shaped structure [26]. Nonetheless, there is little literature discussion on the combination of high-efficient transmission and broadband backward-scattering reduction in different frequency bands by a single structure. Moreover, manipulating electromagnetic responses separately in different frequency bands plays a crucial role in the modern communication system. The frequency selective surface (FSS) which has high transmission efficiency in the passband as well as low radar cross section (RCS) outside the passband is highly desired in many applications, especially in radomes.

The conventional approaches to realize the low RCS band-pass FSS require a large number of lumped resistances to dissipate the electromagnetic energy. In this paper, a novel FSPC is introduced and designed to reduce the back-ward scattering out of the passband as well as acquire high-efficient transmission in the passband. The structure is composed of a fishbone polarization conversion layer backed by a multi-layer FSS with the angle cut on the first layer of the FSS. The multi-layer FSS with the angle cut on the first layer is used to provide a high-efficiency transmission window and a polarization conversion band on the high-frequency side of the passband. While the top fishbone layer is used to achieve a polarization conversion band on the low-frequency side of the passband. The simulation results show that the 1 dB transmission window can be obtained from 8.5 GHz to 11 GHz and the bandwidth of co-polarized reflection below -10 dB is up to 92% from 5.6 GHz to 15.13 GHz. Moreover, the proposed device can achieve two polarization conversion bands (5.66-6.9 GHz and 12.8-15.2GHz) with the polarization conversion ratio over 90%. It is verified by simulation that the backward scattering can be reduced effectively in a wide frequency range from 4GHz to 16GHz when the unit cell structures were arranged in a checkerboard-like distribution. Finally, a prototype of the FSPC consists of 20×20 unit cells are fabricated and measured, the measured results agree well with the simulation ones. Furthermore, the proposed structure is polarization-insensitive and can be applied to many practical situations, for instance, serving as radomes to enable a high-efficient in-band transmission with low-backward scattering for the out-of-band signals.

## 2. Modeling, simulation, and discussion of the FSPC

We start from a multi-layer FSS displayed in Fig. 1(a), which is a combination of two identical square patches on the top and bottom layers respectively, coupled through a square slot in the middle layer. These three metal layers are printed on two identical F4B substrates ($\varepsilon_r = 2.2, tan\delta = 0.002$) with a thickness of $d_1$=3 mm. The period of the three-layer FSS is $p$=15mm and the other geometrical dimensions are $l_1$=7.5mm, $l_2$=6.3mm, and $w$=0.15mm. The full-wave simulation of the bandpass FSS is carried out by CST Microwave Studio and the simulation results are shown in Fig. 1(b). In the following, we define $r_{xy} = E_x^{Ref}/E_y^{Inc}, r_{yy} = E_y^{Ref}/E_y^{Inc}$ as the reflection ratio of $y$-to-$x$ and $y$-to-$y$ polarization conversions and $t_{xy} = E_x^{Trans}/E_y^{Inc}, t_{yy} = E_y^{Trans}/E_y^{Inc}$ as the transmission ratio of $y$-to-$x$ and $y$-to-$y$ polarization conversions, where $E_y^{Inc}$ represents the electric field of the $y$-polarized incident EM wave, and $E_x^{Ref}$, $E_y^{Ref}$, $E_x^{Trans}$ and $E_y^{Trans}$ represent the electric fields of the $x$-polarized, $y$-polarized reflected EM waves and the $x$-polarized, $y$-polarized transmitted EM waves, respectively. The reflection polarization ratio (RPCR) and the transmission polarization conversion ratio (TPCR) of the proposed FSPC are defined as $RPCR = |r_{xy}|^2 / (|r_{xy}|^2 + |r_{yy}|^2 + |t_{xy}|^2 + |t_{yy}|^2)$ and $TPCR = |t_{xy}|^2 / (|r_{xy}|^2 + |r_{yy}|^2 + |t_{xy}|^2 + |t_{yy}|^2)$ respectively. In contrast with the conventional reflective polarization converter backed by a metal ground, the proposed FSPC is backed by an FSS and the transmission coefficient is not zero. Thus, the transmission ratio of $y$-to-$x$ and $y$-to-$y$ polarization conversions should be taken into account when we calculate the RPCR and TPCR.

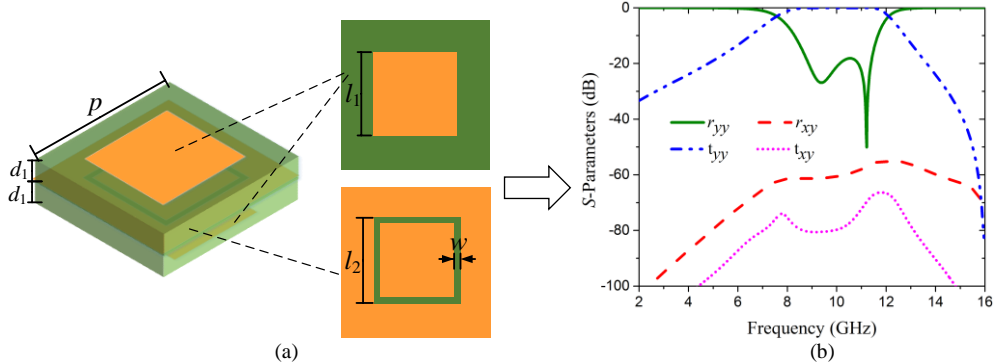

(a)      (b)

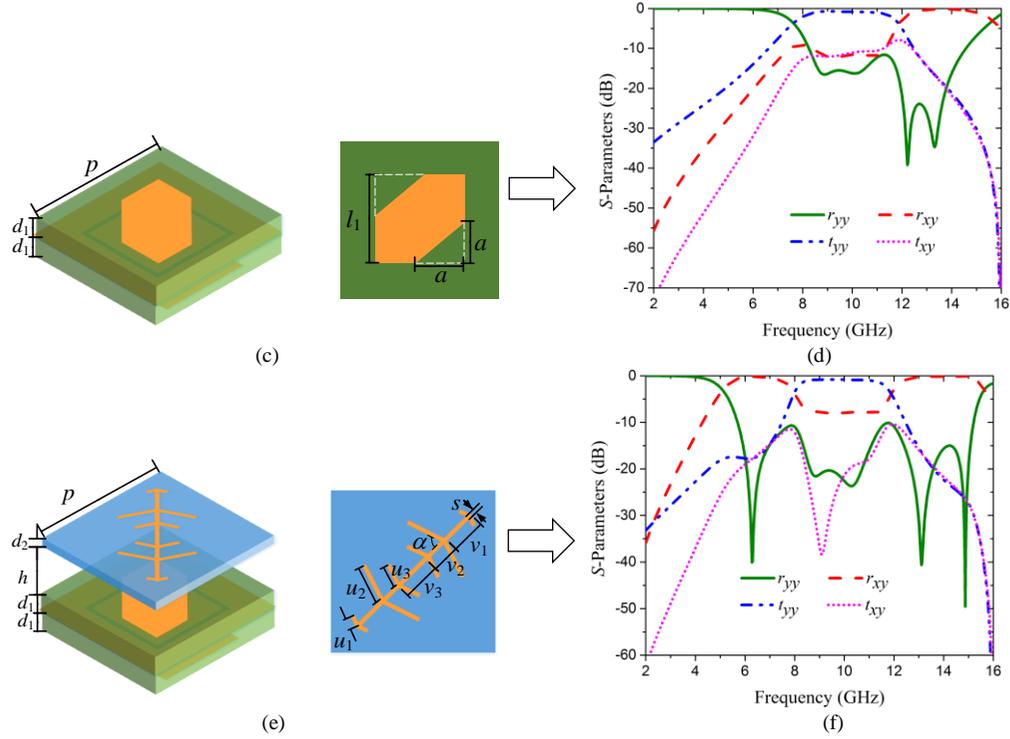

Fig. 1. The scheme, evolution, and simulation from a multi-layer FSS to an FSPC. (a) The unit cell structure of the multi-layer FSS. (b) The simulation results of the FSS. (c) The unit cell of the FSS with two corners cut at the top layer. (d) The simulation results of the FSS with two corners cut at the top layer. (e) The unit cell structure of the FSPC covered by a fishbone layer. (f) The simulation results of the FSPC covered by a fishbone layer.

As shown in Fig. 1(b), the FSS shown in Fig. 1(a) can achieve a broadband co-polarized transmission band from 8GHz to 11.8GHz with the insertion loss less than 1dB and both cross-polarized reflection and cross-polarized transmission are small enough to be ignored. However, when the two corners located on the diagonal line are cut off with the length of $a$=3.5mm (shown in Fig. 2(c)), a reflected polarization conversion band occurs from 12.45GHz to 15GHz with the cross-polarized reflection coefficient $r_{xy} \geq -1$dB (shown in Fig. 2(d)). With a more detailed view of Fig. 2(d), it can be found that the co-polarized transmission band (8.5-11GHz) becomes a little narrower compared with the passband (8-11.8GHz) of the FSS in Fig. 1(a), which is attribute to the increase of $r_{xy}$ and $t_{xy}$ at the passband due to the introduction of anisotropic angle-cut structure. It should be stressed that although the $r_{xy}$ and $t_{xy}$ in Fig. 1(d) increased slightly compared with that in Fig. 1(b), the co-polarized transmission coefficient is larger than -1dB in the passband from 8.5GHz to 11GHz. Moreover, the co-polarized reflection coefficient $r_{yy}$ is less than -10dB from 8.2GHz to 14.6GHz and the small co-polarized reflection coefficient out of the passband is exactly what we desired because it can be utilized to enhance the co-polarized stealth effect and reduce the backward scattering by checkerboard-like distribution design. To acquire the steal performance on the lower frequency range of the passband, a fishbone layer displayed in Fig. 2(e) is designed and covered the structure in Fig. 2(c) with a distance $h$=3.5mm. The substrate of the fishbone layer is the same F4B as the FSS layer with a thickness of $d_2$=0.25mm. The other physical parameters of the fishbone unit cell are summarized in Table 1.

Table 1. Physical Parameters of the Fishbone Unit Cell

| Parameters | $u_1$ | $u_2$ | $u_3$ | $v_1$ | $v_2$ | $v_3$ | $s$ | |
|---|---|---|---|---|---|---|---|---|
| Values (mm) | 0.7 | 3.5 | 1.8 | 2.7 | 1.57 | 3.16 | 0.4 | 72° |

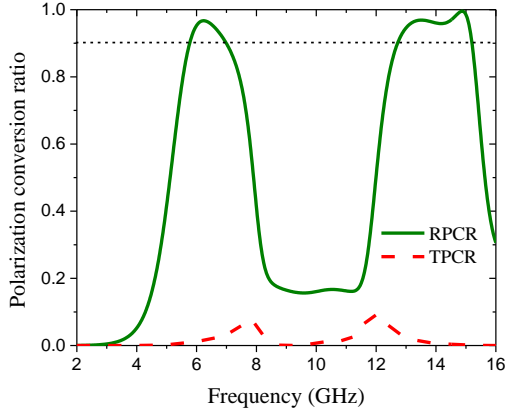

Fig. 2. The calculated RPCR and TPCR of the proposed FSPC.

The final simulation results of the proposed FSPC are shown in Fig. 1(f), the results indicate that the passband and the high-frequency cross-polarized reflection band remain the same as in Fig. 1(d) and there is an additional low-frequency cross-polarized reflection band from 5.5GHz to 7.5GHz with $r_{xy} \geq -1\text{dB}$. What's more, the proposed FSPC can achieve the outstanding co-polarized stealth performance in a wide frequency range from 5.7GHz to 15.2GHz with the co-polarized reflection coefficient $r_{yy} < -10dB$. To gain more insights, we calculate the RPCR and TPCR in Fig. 2, and the results show that the RPCR of the FSPC reaches up to 90% from 5.7GHz to 7GHz and 12.7GHz to 15.2GHz, while the TPCR is negligibly small compared with the RPCR, therefore a high co-polarized transmission coefficient can be achieved in the passband.

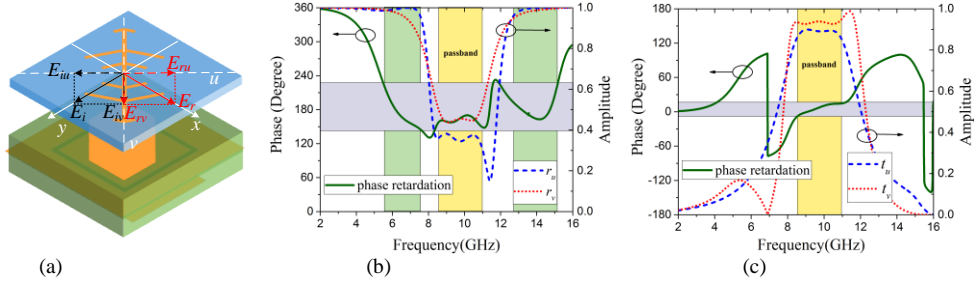

Fig. 3. (a) Intuitive scheme of the FSPC. (b) The reflection amplitudes and phase retardation with the incident electric field along *u*- and *v*-axes, respectively. (c) The transmitted amplitudes and phase retardation with the incident electric field along *u*- and *v*-axes, respectively.

To further analyze the polarization conversion response of the FSPC, we analyze the working principle by decomposing the incident wave $\vec{E}_i$ into two orthogonal directions $\vec{E}_u$ and $\vec{E}_v$, as illustrated in Fig. 3(a). Thus the incident, reflected and transmitted wave can be expressed as $\vec{E}_i = \vec{u}E_u e^{j\phi} + \vec{v}E_v e^{j\phi}$, $\vec{E}_r = \vec{u}r_u E_u e^{j(\phi+\phi_{ru})} + \vec{v}r_v E_v e^{j(\phi+\phi_{rv})}$ $\vec{E}_t = \vec{u}t_u E_u e^{j(\phi+\phi_{tu})} + \vec{v}t_v E_v e^{j(\phi+\phi_{tv})}$, where $r_u$ and $r_v$ ($t_u$ and $t_v$) are the reflection coefficients (transmitted coefficients) along the *u*-axis and *v*-axis, $\phi_{ru}$ and $\phi_{rv}$ ($\phi_{tu}$ and $\phi_{tv}$) are the reflected phases (transmitted phases) along the *u*-axis and *v*-axis. The reflection and transmission response with the *u*-and *v*-polarized incident waves are plotted in Fig. 3(b) and (c) respectively. It can be seen from Fig. 3(b) that the reflection magnitudes ($r_u$ and $r_v$) are nearly equal to one and the phase difference ($\Delta\phi = |\phi_u - \phi_v|$) is roughly 180° at two reflected polarization conversion bands, which leads to the synthetic field $\vec{E}_r$ along the *x*-direction. Thus orthogonal polarization conversion at two reflected polarization conversion bands can be achieved. At the passband, although the phase difference is roughly 180°, the reflection magnitudes are very small, therefore, the RPCR is negligibly small. Furthermore, the transmission magnitudes are nearly equal to one and the phase difference ($\Delta\phi = |\phi_u - \phi_v|$) is roughly 0° at passband which can be seen in Fig. 3(c), namely, the polarization state of the transmitted wave remains unchanged and the incident wave is transmitted with low insertion loss.

## 3. Application of the proposed FSPC to broadband RCS reduction

One potential application of the proposed FSPC is obtained after arranging it in a checkerboard distribution. The RCS reduction of a checkerboard 1-bit metasurface [27], compared with that of the same size PEC, can be approximated by

$$\text{RCSReduction (dB)} = 10\log_{10}\left[\frac{\lim_{R\to\infty} 4\pi R^2 \frac{|\vec{E}_{ry}|^2}{|\vec{E}_{iy}|^2}}{\lim_{R\to\infty} 4\pi R^2 |1|^2}\right]$$

$$= r_{yy}\text{(dB)} \quad (1)$$
$$= 10\log_{10}[1-\text{PCR}]$$

From (1), we can conclude that in the polarization conversion bands, a 10dB backward RCS reduction can be achieved when PCR is greater than 0.9. While in the passband, due to most EM wave can pass through the proposed FSPC, the backward RCS is reduced a lot compared with the same size PEC. Namely, the proposed FSPC can obtain an obvious RCS reduction in a wide frequency range. The low RCS FSPC surface is illustrated in Fig. 4(a), which is a combination of 1-bit coding metasurface using a checkerboard arrangement. The "0" and "1" supercell, which are both made up of 6×6 unit cells, are in orthogonal orientations with each other to obtain 180º phase difference. Fig. 4(b) depicts the simulation monostatic RCS of the proposed checkerboard low RCS surface and the same size PEC. It is observed that an obvious RCS reduction is expected over more than 122% of the frequency bandwidth (from 4.5GHz to 16GHz).

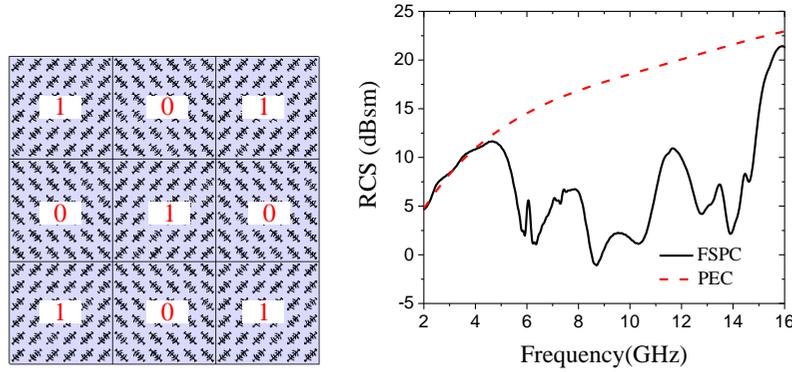

Fig. 4. (a) Structure of the checkerboard low RCS surface composed of the proposed FSPC. (b) The monostatic RCS contrast between the checkerboard low RCS surface and the same size PEC.

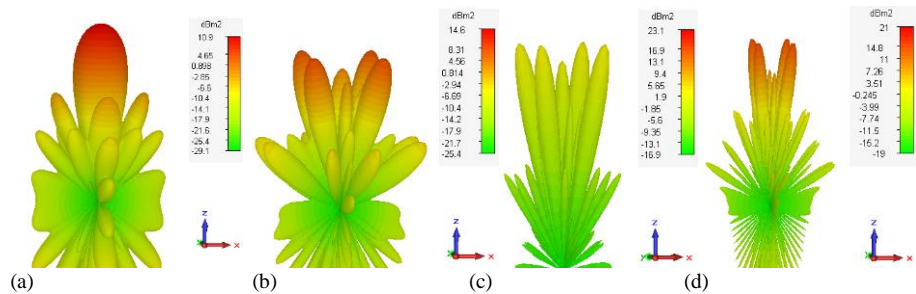

Fig. 5. Three-dimensional scattering patterns of the checkerboard low RCS surface at (a) 4GHz, (b) 6.5GHz, (c) 10GHz, and (d) 15GHz.

To gain more detailed insights, the three-dimensional scattering patterns under normal incidence of the checkerboard low RCS surface at 4GHz (co-polarized reflection band), 6.5GHz (lower cross-polarized reflection band), 10GHz (transmission band), and 15GHz (upper cross-polarized reflection band), which prove the frequency selective scattering capability, are illustrated in Fig. 5. At 4GHz, as shown in Fig. 5(a), the incident EM wave is reflected without transforming the polarization, which results in the strong backward scattering. At the reflected polarization conversion bands (Figs. 5(b) and 5(d)), the reflected EM wave energy is scattered around which can reduce the backward scattering. At the passbands (Fig. 5(c)), the incident EM wave can transmit through the surface, thus the total RCS is lower than the same size PEC. It should be mentioned that since the phase difference between "0" and "1" is 180°, we can use the two elements to form a 1-bit coding metasurface with the random distribution [17] to scatter the incident wave in a diffusion-like way, to reduce the bistatic RCS.

## 4. Experimental verification

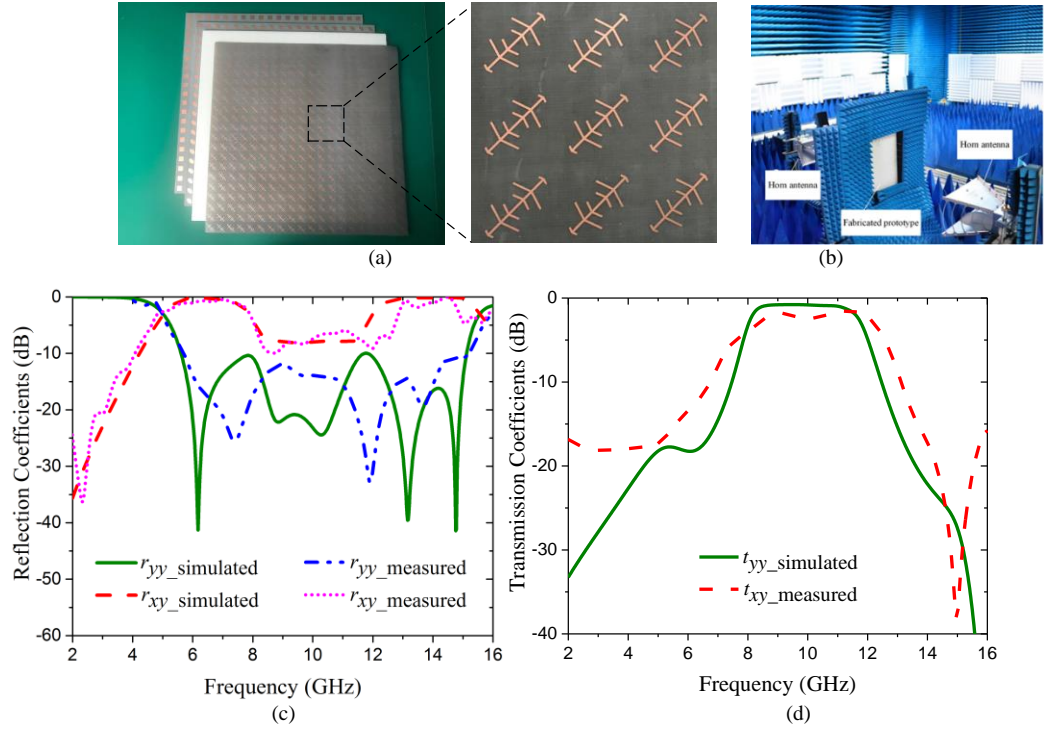

Fig. 6. (a) The fabricated prototype of the proposed FSPC, (b) the measurement setup of the prototype, (c) measured and simulated reflection coefficients under normal incidence, and (d) measured and simulated transmission coefficients under normal incidence.

To validate the performance of the proposed FSPC, the prototype with a size of 300mm×300mm (20×20 unit cell) was fabricated as exhibited in Fig. 6(a). Moreover, the fabricated prototype was measured in the anechoic chamber as shown in Fig. 6(b), which contains two standard horn antennas connected to the vector network analyzer (N5245A). The air space to separate the first fishbone layer and the back FSS layer is realized by a 3.5mm thick PIM with $\varepsilon_r = 1.05$, $tan\delta = 0.005$. The three substrates and the PMI layer are stuck together with glue in the right order. The measured results of the proposed FSPC under the normal incidence are presented in Figs. 6(c) and 6(d), which can be observed that the measured and simulated results agree reasonably. It is demonstrated in Fig. 6(c) that the measured cross-polarized reflection coefficient ($r_{xy}$) is higher than -1.5dB in two frequency ranges (5.5-7.65GHz and 12.9-14.7GHz), and the co-polarized reflection coefficient ($r_{yy}$) is lower than -10dB from 5.7GHz to 15.1GHz, which means a high-efficiency polarization conversion and the outstanding co-polarization stealth capability. And the measured transmission results plotted in Fig. 6(d) shows that the measured insertion loss is less than 2.5dB from 8.5GHz to 12.1GHz, which approves that the proposed FSPC can obtain a broadband high-efficiency passband. The small discrepancy between the experiment and simulation may be caused by the manufacture and measurement tolerances.

## 5. Conclusion

In summary, a novel FSPC combing reflection polarization converter and FSS together was designed and measured. Both the full-wave simulation and measurement of the fabricated prototype verify that the proposed FSPC can achieve a broadband passband with low insertion loss and two high-efficiency reflection polarization conversion bands on both sides of the passband. Furthermore, we use the two orthogonal orientations of the FSPC unit cells as the "0" and "1" elements to form a 1-bit surface with a checkerboard distribution to reduce the backward RCS. The simulation results of the checkerboard surface show that an obvious RCS reduction can be achieved in a wide frequency range from 4.5GHz to 16GHz. Providing high-efficiency transmission in the predesigned band, the FSPC is nearly "invisible" in the entire working frequency band to the microwave detections with a good performance. This unique feature may be applicable for various applications both in the microwave and terahertz bands. For example, working as the stealth radomes to enable a high-efficient in-band transmission with out-of-band low-backward scattering.